\documentclass[aps,preprint,amssymb, amsmath,prd]{revtex4}
\usepackage{bm}
\usepackage[usenames,dvipsnames]{color}
\usepackage{graphicx}
\usepackage{ulem}

\newcommand{\edm}{\end{displaymath}}
\newcommand{\beq}{\begin{equation}}
\newcommand{\eeq}{\end{equation}}
\newcommand{\beqa}{\begin{eqnarray}}
\newcommand{\eeqa}{\end{eqnarray}}

\newcommand{\edo}{\end{document}}

\newcommand{\bit}{\begin{itemize}}
\newcommand{\eit}{\end{itemize}}
\newcommand{\ben}{\begin{enumerate}}
\newcommand{\een}{\end{enumerate}}

\newcommand{\rar}{\rightarrow}

\newcommand{\bfP}{{\bf P}}
\newcommand{\bfp}{{\bf p}}

\newcommand{\bfB}{{\bf B}}

\newcommand{\bfE}{{\bf E}}

\newcommand{\bfF}{{\bf F}}
\newcommand{\bfg}{{\bf g}}
\newcommand{\bfG}{{\bf G}}
\newcommand{\bfL}{{\bf L}}
\newcommand{\bfR}{{\bf R}}

\newcommand{\bfa}{{\bf a}}

\newcommand{\bfl}{{\bf l}}
\newcommand{\bfn}{{\bf n}}

\newcommand{\bfbe}{{\mbox{\boldmath$\beta$}}}

\newcommand{\bfr}{{\bf r}}

\newcommand{\bfv}{{\bf v}}

\newcommand{\Rcb}{\mbox{\boldmath${\cal R}$}}
\newcommand{\veps}{{\varepsilon}}

\begin{document}

\title{Basic observables for the accelerated  electron and its  field}

\author{Mihai Dondera}
\email{mihai.dondera@fizica.unibuc.ro}
\affiliation{Department of Physics  and Centre for Advanced Quantum Physics, University of Bucharest, MG-11, Bucharest-M\u agurele, 077125  Romania}
\date{\today}

\begin{abstract}
We revisit in the framework of the classical theory the problem of the accelerated motion of an electron, taking into account the effect of the radiation emission. 
We present results for the momentum and energy of the electromagnetic field of an accelerated electron  {for a spatial region  excluding a vicinity of the electron} and a procedure to compensate their singularities in the limit of the point electron.  { From them we  infer expressions  for} the observables  momentum and  energy of the electron. They lead, together with those corresponding to the emitted radiation,  to an equation of motion of the electron  that  coincides,  in the case of  an external electromagnetic field,  with the Lorentz-Dirac equation. {
Based on the results for the linear momentum and using the same calculation method, we obtain the equations corresponding to the angular momentum of the electron and its field.} While the formalism used is not wholly manifestly covariant, arguments based on relativistic covariance are invoked { and used} at appropriate places, where they play an essential role.
\end{abstract}

\maketitle

\section{Introduction}

\label{intr}

The collision processes of  high energy electrons with other charged particles or with  intense  laser pulses can be strongly influenced by the emission of radiation accompanying these processes. The experimental observation of this influence or reaction,  usually called {\it radiation reaction} (RR),  in the case of electrons colliding with laser pulses,  as well as of other strong fields effects and processes, became  possible  nowadays  due to the remarkable progresses registered by the laser technology -- see \cite{adpk} for a recent review of high-energy processes in extremely intense laser fields. Two  very recent papers \cite{poder,cole} present and analyze experimental results proving for the first time a  strong radiation reaction regime in the interaction of relativistic electrons with ultra-high intensity laser pulses. They illustrate the actual limitations for observing  RR and  the inherent difficulties to validate the theoretical models  for RR. {While non-negligible  quantum effects are predicted for conditions of the quoted experiments, the data obtained seem to not allow a clear  discrimination between classical and quantum models -- see also the viewpoint expressed in \cite{macchi}}. 

The above mentioned actual context and the future experiments planned  for higher intensities of laser pulses at other  laser facilities (as, for example, at ELI-NP \cite{turcu}) indicate the need for a better  description of RR, in both classical and quantum theory. { In particular,  one may appreciate that a progress in  the classical theory of RR   (the only one considered in this paper) in   understanding  the basic observables  of the electron and its field,  would lead to a more consistent description of the reaction effect. This potential progress should  be also  beneficial  for the formulation of  a more accurate quantum description of RR and for an improved understanding of the quantum-classical correspondence relation.}

We recall that the  description of the accelerated motion of an electron with the inclusion of RR {-- see \cite{hammond, burt} for qualified introductions on the subject, discussions of the main equations used for RR and for the relevant references --} presents considerable difficulties (no matters  the cause of acceleration,  the presence of an external field or the interaction with other particles), related to the treatment   of the electromagnetic field of the moving electron  - the evolution of this   field depends on the motion of the electron and, in turn, influences this motion. { Rather interesting and somehow unexpected,    the problem of finding  the {\it reaction} of the electromagnetic field on the motion of an electron accelerated by an external field admits a closed solution (recalled below).}
Most frequently, the approaches used to solve this problem   concentrate on the forces which determine the motion of the electron. Besides  the external force acting on the electron one  has to take into account the apparent force related   to  the  momentum transported by the electromagnetic field of the moving electron. This reaction force is  called {\it radiation reaction} force or {\it radiation damping} (or   {\it friction}) force, terminology used even if one cannot fully explain the origin of the reaction on the base of the emitted radiation. 
In fact, two types of electromagnetic reaction forces can be identified, one of them is directly  connected with the radiation and vanishes in the non-relativistic limit, the other being in the same limit proportional to the time derivative of the electron acceleration  and being termed sometimes  Abraham (or Abraham-Lorentz) force or {\it self-force}.

The existing calculations of the electromagnetic reactions consider the electron either as an extended charged particle of finite size with a corresponding charge distribution (the Lorentz model) or as a point charge. The Lorentz model of the electron was fully abandoned in  \cite{dirac}, where  a formalism based on retarded and advanced potentials was build for a point charged particle, culminating  with the derivation of  the equation {of motion} of the electron.
This  equation  coincides with that emerging approximately from the  Lorentz model and it is usually termed Lorentz-Abraham-Dirac (LAD) or simply Lorentz-Dirac equation. {The problem of the uniqueness of the motion equation was analyzed in \cite{bhabha} and it was shown that for a point particle in electromagnetic field the conservation of the angular momentum, not implied in general by this equation, has to be imposed distinctly. This leads to supplementary constraints on the form of the  equation, which is unique and coincides with LAD equation if one requires also to not contain  derivatives of the electron velocity of an order higher than the second one.}
 {Another derivation of LAD equation}, based on a decomposition of the Maxwell tensor and not relying on the advanced fields, was presented in \cite{titbo}.

The issue related to the order  of LAD equation (third order with respect to time)  and its unphysical consequences is overcome if in place of it one works with Landau-Lifshitz (LL)  equation \cite{ll}. This second order equation, being an approximation of LAD equation, was questioned  about its validity range, investigation presenting an increasing interest for the actual context. { Alternative theories to LAD and LL equations where described in the literature -- see \cite{burt} and references therein. We mention here Ref. \cite{soko},  which attracted some interest  in the last years, in particular for the possibility to regard the relation of the velocity and the momentum of the electron as being a Lorentz transformation induced by the external electromagnetic field \cite{capdes}.  The   set of equations proposed in \cite{soko}, replacing approximately LAD equation, and  equivalent  with LL equation but numerically simpler  \cite{capdes}, is based on the idea that the 4-vectors velocity and momentum  of an accelerated electron might be introduced such they are not-collinear, and on supplementary assumptions, in particular that of the validity of the momentum-energy Einstein relation.  }
 
The approach  employed in this paper to investigate the electromagnetic reaction is based on the examination of the fundamental observables of the compound system (electron and electromagnetic field), followed then  by the analysis of their implications. 
In place of directly targeting the  equations of motion,  we focus  on the problem of finding the expressions of the {
momentum,  energy and angular momentum} of an electron accelerated by an external force. 
We adhere to the assumptions  of \cite{dirac} - the electron is a point charged particle, the Maxwell equations are supposed  valid everywhere  and, consequently, the fields are singular at  the electron position. 
{
In Sect. \ref{mae}} 
 we first describe  the results obtained for the momentum and energy of the electromagnetic field of an accelerated electron. The method we use  to calculate them, presented in Appendix, allows to clearly separate the terms corresponding to the emitted radiation from those bound to the electron.  In order to treat  the singularities (associated with the zero size of the electron) of the latter terms  and their  partial covariance,   we adopt a simple  compensation procedure then we infer the expressions of the electron observables, analyze  their main features and follow some implications.  From the formulas for the  momentum and energy of the electron and its electromagnetic field we  simply re-derive  the electromagnetic reaction forces and formulate the equations of motion of the electron, finally re-obtaining the LAD  equation.

{
In Sect. \ref{amo} we refer to another important observable, the angular momentum. The starting point is the angular momentum of the electromagnetic field of the accelerated electron. Its calculation is performed by the {method used} in Sect. \ref{mae} and is also  described in Appendix. Adding  to it the angular momentum of the bare electron, then considering the limit of the point electron we finally 
obtain the expression of the total angular momentum. In the last part of Sect. \ref{amo} we examine the balance of the angular momentum, the generalization of the corresponding theorem and some consequences of it.
}

\section{Momentum and energy}

\label{mae}

For the beginning it is convenient to review some basic relations of the relativistic kinematics   for the case of an electron in {\it arbitrary} motion. We denote by $\bfr_e(t)$ the position of the electron as function of the time $t$ and by $x^\mu\equiv (ct,\bfr_e)$  the corresponding 4-vector of the position. With $\bfv\equiv \dot{\bfr}_e\equiv d\bfr_e/dt$ the (usual) velocity of the electron, the Lorentz factor is 
\beq
\label{lorfac}
\gamma\equiv\frac{1}{\sqrt{1-\bfbe^2}}, \quad  \bfbe\equiv\frac{\bfv}{c}.
\eeq
Differentiating successively  the 4-position to $\tau$,  the electron proper time, defined by
\beq
d\tau=\sqrt{1-\bfbe^2}\,dt=\frac{dt}{\gamma}, 
\eeq
one obtain the 4-vectors velocity $u^\mu\equiv (u^0,{\bf u})$ and acceleration  $a^\mu\equiv (a^0,\bfa)$. They are given by
\beq
\label{4velo}
u^\mu=\frac{dx^\mu}{d\tau}= (\gamma c,\gamma\bfv)=c\,(\gamma,\gamma \bfbe)
\eeq 
and
\beq
\label{4acc}
a^\mu=\frac{du^\mu}{d\tau}=c \,\left(\gamma\, \frac{d\gamma}{dt}, \gamma\, \frac{d}{dt}(\gamma\bfbe)\right)=
c\,\gamma^2 \,\left({\gamma}^2(\bfbe\cdot\dot{\bfbe}),\dot{\bfbe}+\gamma^2(\bfbe\cdot\dot{\bfbe})\bfbe\right).
\eeq
The  4-scalar  
\beq
\label{a2a02}
\bfa^2-(a^0)^2={c^2}\gamma^4\left[ {\dot{\bfbe}}^2+\gamma^2(\bfbe\cdot\dot{\bfbe})^2 \right]=c\gamma^2\dot{\bfbe}\cdot\bfa
\eeq
presents a special interest in the following. Another relation  implied  by Eqs. (\ref{4velo}) and (\ref{4acc}),
\beq
\label{a0ba}
 a^0=\bfbe\cdot \bfa, 
\eeq
{equivalent to the orthogonality of 4-velocity and 4-acceleration,  proves also to be useful.}

 We next recall that a {\it free} electron, moving with  constant velocity $\bfv$, has a linear momentum $\bfp_e$ and a total energy $W_e$, given by  
\beq
\label{pwfe}
\bfp_e=m\gamma\bfv,\quad W_e=mc^2\gamma,
\eeq
where $m$ is the electron mass. These quantities 
satisfy the well-known equation  
 \beq
 \label{wp}
 W_e^2=m^2c^4+c^2\bfp_e^2,
 \eeq
relating the particle energy to the magnitude of its linear momentum. The momentum and energy (divided by $c$) of the particle  { are the components  of }the 4-momentum
\beq
p_e^\mu=(W_e/c,\bfp_e),
\eeq
proportional to the 4-velocity (\ref{4velo}), $p_e^\mu=mu^\mu$.

We investigate in the following  how the above  relations for momentum and energy of a free electron are modified in the case of an electron  moving in the field of an external force  $\bfF$.    These modifications are expected since, due to the fact that an accelerated electron  emits electromagnetic radiation, its motion does not coincide with that of  a neutral particle of the same mass which would be exposed to the same force $\bfF$.

{ For finding the relations replacing Eq.  (\ref{pwfe}) in the case of an {\it accelerated} electron,  we  first take in consideration that the radiation emitted by the electron during its motion up to the  actual time $t$,  carries momentum and energy and denote these quantities by $\bfp_{rad}(t)$ and $W_{rad}(t)$. For the  momentum  and  energy of the electron  we use  the same notations as in the case of the  free electron, $\bfp_e(t)$ and $W_e(t)$ (they are now functions of time). 
If the electron was accelerated in the past but is {\it free} at the time  $t$, {\it i.e.} $\dot{\bfbe}(t)=0$, $\bfp_e(t)$ and $W_e(t)$ are again given by Eq. (\ref{pwfe}) and the total momentum $\bfP$ and the total energy $W$ of the compound system (electron and radiation) can be expressed as follows 
\beq
\label{pwt0}
\bfP(t)=\bfp_e(t)+\bfp_{rad}(t),\quad W(t)=W_e(t)+W_{rad}(t).
\eeq 
 Two types of contributions  shall be supposed below as being  included in $\bfp_e(t)$ and $W_e(t)$, of the own field of the electron (field not-leaving the electron as radiation), and of  the "bare" electron, regarded as the other facet of the entity "electron". With further interpretation, these contributions, which  combine to reproduce the results (\ref{pwfe}) in the case of the free electron, are such Eq.  (\ref{pwt0}) is verified for the accelerated electron too (by definition, no contributions are included in $W_e$ corresponding to  the interaction with the external field).
}

{In order to reach the above announced objective} we examine first the problem  of the {\it total} momentum and energy of the  electromagnetic field  accompanying  an  accelerated electron. For their calculation  we use the general { expressions given  by} classical electrodynamics for the momentum density $\bfg$ and energy density $u$  of the  field 
\beq
\label{gu}
\bfg=\varepsilon_0\,\bfE\times \bfB, \quad u=\frac{\varepsilon_0}{2}\left(\bfE^2+c^2\bfB^2\right),
\eeq
together with the expressions of the electric field $\bfE$ and magnetic field $\bfB$ generated by  an electron in its motion, the so-called Li\'{e}nard-Wiechert fields. At an observation point $\bfr$ and for the {\it actual} time moment $t$,  these are given by \cite{jack}
\beq
\label{ebfields}
\bfE(\bfr,t)=\frac{e}{4\pi\varepsilon_0}\frac{1-\beta^2}{R^2\varkappa^3}\,(\bfn-\bfbe)+\frac{e}{4\pi\varepsilon_0 c}\frac{1}{R\varkappa^3}\,\bfn\times\left[(\bfn-\bfbe)\times\dot{\bfbe}\right]
, \quad \bfB(\bfr,t)=\frac{1}{c}\,\bfn\times \bfE(\bfr,t),
\eeq
where $\bfn$ is the unit vector along $\bfR\equiv \bfr-\bfr_e$ and $\varkappa\equiv 1-\bfn\cdot\bfbe$. 
The expressions of the fields  have to be evaluated at the {\it retarded}  {{ time moment $t'$,  verifying the equation  
\beq
\label{tret}
t'=t-R(t')/c, \quad R(t')=|\bfr-\bfr_e(t')|,
\eeq
and uniquely determined by $\bfr$ and $t$ (for a given motion $\bfr_e(t)$ of the electron). }

\begin{figure}
\includegraphics[width=0.5\columnwidth]{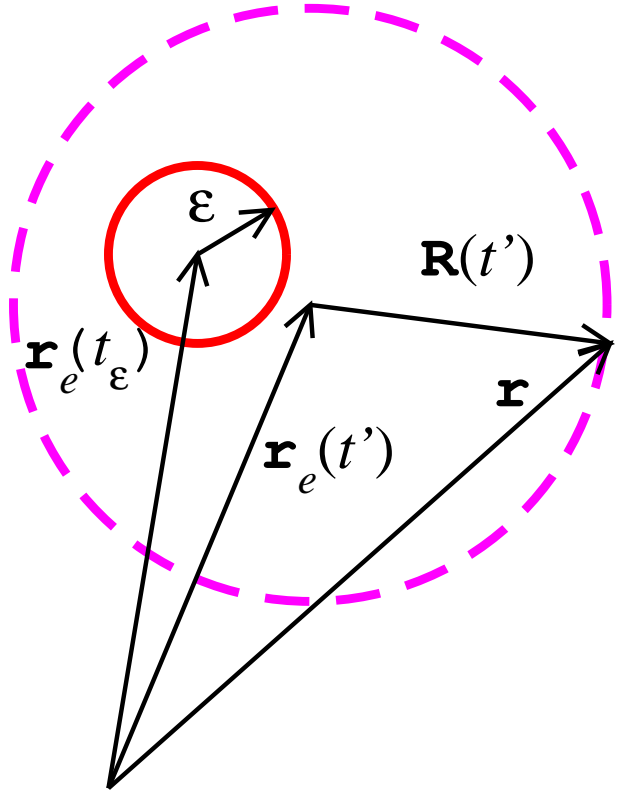}
\caption{(Color online) The region $\cal D$ considered in the paper for integration is the exterior of the sphere  (represented by the   solid line circle) of radius $\veps$.   The vector $\bfr$ describes an arbitrary position in $\cal D$, while $\bfr_e(t')$ is the electron  position at the retarded time moment $t'$ (anterior to $t_{\veps}$), verifying Eq. (\ref{tret}). }
\label{geo}
\end{figure}

Like the fields, the densities (\ref{gu}) are singular at $\bfr=\bfr_e(t)$, the actual electron position. These singularities are not-integrable in general, the integration of the densities  (\ref{gu}) over { a region containing the singularity}  leading to infinite values for total momentum and total energy. {This difficulty can be postponed } calculating first  the  momentum and energy (at a given time $t$) of the field in a spatial region $\cal D$  taken to be the whole space minus a small region around the particle, then examining  the limit where that small region reduces to the point $\bfr_e(t)$.
The appropriate choice for  ${\cal D}$ is that of a region 
outside a sphere of radius $\veps$ (see Figure \ref{geo}), centered on the position $\bfr_e(t_{\veps})$ of the electron at the time moment $t_{\veps}\equiv t-{\veps}/c$. {
 With this choice  the spatial integration on $\cal D$ 
can be performed changing the integration variables from the coordinates $\bfr$   to the spherical coordinates of $\bfR(t')=\bfr-\bfr_e(t')$. The calculation, presented in the Appendix, leads  to the following results for the  momentum 
\beq
\label{glam2}
\bfG(\veps,t)=\frac{2}{3}\frac{e_0^2}{c}\frac{\gamma^2(t_{\veps})\bfbe(t_{\veps})}{\veps}    +\frac{2}{3}\frac{e_0^2}{c^2}\int_{-\infty}^{t_{\veps}}
\gamma^4\left[ {\dot{\bfbe}}^2+\gamma^2(\bfbe\cdot\dot{\bfbe})^2 \right] \bfbe\,dt',
\eeq
and for the energy 
\beq
\label{wd}
W(\veps,t)=\frac{2}{3}\,{e_0^2}\frac{\gamma^2(t_{\veps})-1/4}{\veps}+\frac{2}{3}\frac{e_0^2}{c}\int_{-\infty}^{t_{\veps}}
\gamma^4\left[ {\dot{\bfbe}}^2+\gamma^2(\bfbe\cdot\dot{\bfbe})^2 \right]\,dt'
\eeq
of the electromagnetic field in $\cal D$, with the notation $e_0\equiv e/\sqrt{4\pi\veps_0}$. The terms containing the integrals are the contributions of the acceleration fields, the others include  velocity fields and mixed contributions (see the Appendix).

We examine the case where ${\cal D}$ covers the whole space, this meaning  to look to the limit $\veps\rar 0$ (
$t_{\veps}=t-\veps /c\rightarrow t$)  in Eqs. (\ref{glam2}) and (\ref{wd}). 
{ 
The  last terms in these equations do not raise problems  when $\veps\rar 0$ and we identify them with  the momentum 
\beq
\label{prad}
\bfp_{rad}=\frac{2}{3}\frac{e_0^2}{c^2}\int_{-\infty}^{t}
\gamma^4\left[ {\dot{\bfbe}}^2+\gamma^2(\bfbe\cdot\dot{\bfbe})^2 \right] \bfbe\,dt' 
\eeq
and the energy
\beq
\label{wrad}
W_{rad}=\frac{2}{3}\frac{e_0^2}{c}\int_{-\infty}^{t}
\gamma^4\left[ {\dot{\bfbe}}^2+\gamma^2(\bfbe\cdot\dot{\bfbe})^2 \right]\,dt' 
\eeq
of the radiation emitted up to the time $t$. 
{
 The energy radiated by the electron per unit of time, $dW_{rad}/dt$, is the well-known Li\'{e}nard-Larmor  power 
\beq
\label{LP}
{\cal P}_L=\frac{2}{3}\frac{e_0^2}{c}\,
\gamma^4\left[ {\dot{\bfbe}}^2+\gamma^2(\bfbe\cdot\dot{\bfbe})^2 \right]=\frac{2}{3}\frac{e_0^2}{c^3} \gamma^4\left[ \dot{\bfv}^2+\gamma^2(\bfbe\cdot\dot{\bfv})^2 \right],
\eeq
where $\dot{\bfv}=d\bfv/dt$ is the usual 3-vector acceleration.  Being proportional to the 4-scalar (\ref{a2a02}), ${\cal P}_L$ is  a 4-scalar {in its turn}. We remark that the rate of variation of $\bfp_{rad}$ (the time derivative $d\bfp_{rad}/dt$) has the direction of the electron velocity and is equal to the product of the mass radiated per time unit, ${\cal P}_L/c^2$, by the velocity $\bfv=c\bfbe$ of the electron. The negative of this rate ($-d\bfp_{rad}/dt$) appears as a damping force due to radiation (see Eq. (\ref{rare}) from below). 
 The quantities $W_{rad}/c$ and $\bfp_{rad}$ are the components of a 4-vector, the 4-momentum of the emitted radiation.

The other terms entering} in Eqs. (\ref{glam2}) and (\ref{wd}), denoted conveniently 
\beq
\label{pfat0}
{\bfp}_f(\veps,t)\equiv \frac{2}{3}\frac{e_0^2}{c}\frac{\gamma^2(t_{\veps})\bfbe(t_{\veps})}{\veps}, 
\eeq 
and
\beq
\label{wfat0}
 W_f(\veps,t)\equiv \frac{2}{3}\,{e_0^2}\frac{\gamma^2(t_{\veps})-1/4}{\veps},
\eeq 
are, respectively, the momentum and the energy of the  field (in the region $\cal D$) not leaving the electron as electromagnetic radiation.  These terms are bound to the electron and we note that they are singular in the limit $\veps\rar 0$ in the general case. There is an exception, however,  where the singularity does not manifest. This is related to the  momentum of the field of  an electron {\it instantaneously} at rest at the actual time $t$, e.g. $\bfbe(t)=0$. For this case, replacing in Eq. (\ref{pfat0}) $\bfbe(t_{\veps})$ by its  Taylor series expansion $\bfbe(t_{\veps})=\bfbe(t-{\veps}/c)=-({\veps}/c)\,\dot{\bfbe}(t)+{\cal O}(\veps^2)$, it follows that  $\bfp_f(\veps,t)$  has a finite limit for $\veps\rar 0$, {proportional to the (usual) acceleration of the electron,}
\beq
\label{pfrest}
\bfp_f(t)=-\frac{2}{3}\frac{e_0^2}{c^2}\,\dot{\bfbe}(t)=
-\frac{2}{3}\frac{e_0^2}{c^3}\,\dot{\bfv}(t).
\eeq
Its negative rate of changing $(-d\bfp_f/dt)$ coincides with the non-relativistic Abraham force (self-force) [see below Eq. (\ref{selff}) for its relativistic generalization].
In the same limit the field energy can be approximated as 
$
W_f(\veps,t)\approx \frac{1}{2}\,{e_0^2}/{\veps}, 
$ 
and tends to infinity when ${\veps}\rar 0$. 

{In passing, for $\bfv(t)\ne 0$, we may note that in the non-relativistic theory, in the first order of $\bfv/c$ and for a fixed small $\veps$, replacing $\gamma(t_\veps)=1$ in Eqs. (\ref{pfat0}) and (\ref{wfat0}) we can write ${\bfp}_f(\veps,t)=m'\,\bfv$ and $W_f(\veps,t)=m''c^2$, where the masses $m'$ and $m''$ depend on $\veps$ and have  the ratio  $m'/m''=4/3$ - this is a "reincarnation" of the infame "4/3" problem, first met in the case of an extended electron. }

Returning to the general case, one sees that, besides the issue of singularities in the limit ${\veps}\rar 0$, Eqs.  (\ref{pfat0}) and (\ref{wfat0}) display an improper  relativistic behavior.  {This feature is not so  surprising -  as discussed in \cite{jack},  integrating spatially the densities (\ref{gu}) for a general electromagnetic field, the resulting  field energy (divided by $c$) and momentum form a 4-vector if the electromagnetic stress tensor is divergenceless, condition not fulfilled in the presence of the sources (our case).} {One way   to remedy the covariance "defect" is to modify the expressions of momentum and energy of the field 
{
-- see, for example,  Refs. \cite{titbo} and \cite{jack}}. {We don't pursue this thread here since it seems not possible to sustain with convincing arguments that the own field momentum has to be a true observable. }   Another possibility, preferred in the following, is to  admit that the  "bare" electron contributions are such to "fix" both the singularity issue and the covariance property.  
While the nature of these contributions can't be 
elucidated in the context of the present theory, we may assume that they are such to compensate 
   the  above mentioned singularities  in the limit ${\veps}\rar 0$, at the same time leading to correct relativistic  expressions for the total momentum and energy of the electron 
and reproducing in particular the relations (\ref{pwfe}) for the free electron case.

These considerations shall be applied below, after we analyze in detail  Eqs.  (\ref{pfat0}) and (\ref{wfat0}) around $\veps=0$. Their structure becomes transparent} if one replace the functions of $t_{\veps}$  contained in these equations by   their Taylor  series expansions  around the actual time $t$.  { One obtain  the power series (in $\veps$)}
\beq
\label{pfat1}
{\bfp}_f(\veps,t)\equiv \frac{2}{3}\frac{e_0^2}{c}\frac{\gamma^2\bfbe}{\veps}-\frac{2}{3}\frac{e_0^2}{c^2}\frac{d}{dt}(\gamma^2\bfbe)+{\cal O}(\veps), 
\eeq 
and 
\beq
\label{wfat1}
 W_f(\veps,t)\equiv \frac{2}{3}\,{e_0^2}\frac{\gamma^2-1/4}{\veps}
 -\frac{2}{3}\,\frac{e_0^2}{c} \frac{d}{dt}(\gamma^2)+{\cal O}(\veps)
\eeq 
(the time argument $t$ on the right side has been 
 removed) - since we are interested about the limit $\veps \rar 0$, it is sufficient to analyze only the terms explicitly written.}

 The singular terms in the right hand side of Eqs. (\ref{pfat1}) and (\ref{wfat1}), of the order $1/\veps$, coincide with  the momentum and energy of the field in the case of a free electron that would have the velocity $c\bfbe(t)$.  The terms on the second position, finite and independent on $\veps$, do not have (as also the first ones)  proper relativistic behavior. We isolate from them  a 4-vector whose spatial part  reduces to the result (\ref{pfrest}) for $\bfbe(t)\rar 0$. Replacing first the derivative of $\gamma^2\bfbe$ in Eq. (\ref{pfat1}) by the formula for the product $\gamma(\gamma\bfbe)$, one obtains the equivalent form  
\beq
\label{pfat2}
{\bfp}_f(\veps,t)\equiv \frac{2}{3}\frac{e_0^2}{c}\frac{\gamma^2\bfbe}{\veps}- \frac{2}{3}\frac{e_0^2}{c^2}\dot{\gamma}\,\gamma\bfbe+\bfp_S+{\cal O}(\veps),
\eeq 
where the last term  
\beq
\label{peS}
\bfp_S\equiv -\frac{2}{3}\frac{e_0^2}{c^2}\,\gamma\,\frac{d}{dt}(\gamma\bfbe)=-\frac{2}{3}\frac{e_0^2}{c^3}\,\gamma\,\frac{d}{dt}(\gamma\bfv),
\eeq
is the spatial part of a 4-vector proportional to the 4-acceleration (\ref{4acc}). The corresponding time component is 
\beq
\label{peS0}
p_S^0= -\frac{2}{3}\frac{e_0^2}{c^2}\,\gamma\,\frac{d\gamma}{dt},
\eeq
and one sees that the energy 
\beq
W_S\equiv cp_S^0=-\frac{2}{3}\frac{e_0^2}{c}\,\dot{\gamma}\gamma
\eeq
 is one half of the last term in Eq.  (\ref{wfat1}), so we can write
\beq
\label{wfat2}
 W_f(\veps,t)\equiv \frac{2}{3}\,{e_0^2}\frac{\gamma^2-1/4}{\veps}
 -\frac{2}{3}\,\frac{e_0^2}{c}\dot{\gamma}\,\gamma
+W_S + {\cal O}(\veps).
\eeq

{
We note a special feature of the relations (\ref{pfat2}) and (\ref{wfat2}): they describe quantities which depend solely on the state of the electron 
at the {\it present} time moment $t$, in contrast with $\bfp_{rad}$ and $W_{rad}$ from Eqs. (\ref{prad}) and (\ref{wrad}), that depend on the whole past motion of the electron. If these quantities would be fully acceptable, this feature would enable  us to regard them as characterizing  the actual electron state. Since they are not, we are left with an indication that the problem has to be solved at the level of the electron itself. In this order we shift the attention to the other 
facet of the electron, the "bare" electron, and consider its contributions to the momentum and  energy of the electron, denoted,  respectively, by $\overline{\bfp}_e(\veps,t)$ and  $\overline{W}_e(\veps,t)$.  For the  total quantities of the electron we then have
\beq 
\label{pef3}
{\bfp}_e(\veps,t)\equiv \overline{\bfp}_e(\veps,t)+\bfp_f(\veps,t), \quad 
 {W}_e(\veps,t)\equiv \overline{W}_e(\veps,t)+W_f(\veps,t),
\eeq 
and we build  the power series for $\overline{\bfp}_e(\veps,t)$ and $\overline{W}_e(\veps,t)$  such to compensate both the singular and the non-covariant terms in Eqs. (\ref{pfat2}) and (\ref{wfat2}), and  to lead to  correct results for the free electron case. 
We infer that the power series in $\veps$ for   $\overline{\bfp}_e(\veps,t)$ and $\overline{W}_e(\veps,t)$   are, respectively,
\beq
\overline{\bfp}_e(\veps,t)=-\frac{2}{3}\frac{e_0^2}{c}\frac{\gamma^2\bfbe}{\veps}+ \frac{2}{3}\frac{e_0^2}{c^2}\dot{\gamma}\,\gamma\bfbe+mc\gamma\bfbe+{\cal O}(\veps),
\eeq
and
\beq
\overline{W}_e(\veps,t)=
-\frac{2}{3}\,{e_0^2}\frac{\gamma^2-1/4}{\veps}
 +\frac{2}{3}\,\frac{e_0^2}{c}\dot{\gamma}\,\gamma
 +mc^2\gamma+{\cal O}(\veps).
 \eeq
The terms on the first two positions in each equation are, up to the sign, the same as in Eqs. (\ref{pfat2}) and (\ref{wfat2}). On the last positions are included the momentum and energy of a free electron with velocity $c\bfbe(t)$ - this guarantees the correctness of the equations for $\bfp_e$ and $W_e$ in the case the acceleration vanishes.
 
Passing to the limit ${\veps}\rar 0$ in Eqs. (\ref{pef3}), we finally obtain the expressions for the electron momentum, 
\beq
\label{pef2}
\bfp_e=mc\gamma\bfbe+\bfp_S,
\eeq
and its energy, 
\beq
\label{wet}
W_e=mc^2\gamma+W_S 
\eeq
(all the quantities are at the same time $t$). 
The last term in Eq. (\ref{wet}) is the so called Schott  energy (or acceleration energy), $W_S=cp_S^0$. Using Eq. (\ref{peS0}) and the rate $\dot{\gamma}=\gamma^3\,\bfbe\cdot\dot{\bfbe}$ of the Lorentz factor $\gamma$, it can be expressed as 
\beq
\label{wsh}
W_S=-\frac{2}{3}\frac{e_0^2}{c}\,\gamma^4\,\bfbe\cdot\dot{\bfbe}=-\frac{2}{3}\frac{e_0^2}{c^3}\,\gamma^4\,\bfv\cdot\dot{\bfv},
\eeq
relation showing that $W_S$ vanishes when either $\bfv$ or $\dot{\bfv}$ vanishes, or $\bfv$ and $\dot{\bfv}$ are orthogonal.

We interpret Eqs. (\ref{pef2}) and (\ref{wet})  as generalizing the expressions (\ref{pwfe})} of the  momentum and energy of the electron, here taking into account  that the electron is accelerated. 
The supplementary terms, the  momentum $\bfp_S$, named {\it Schott momentum} in the following,   and the energy $W_S$, may contribute solely in the case of non-vanishing electron acceleration. Equation (\ref{pef2}) shows that the momentum of the electron is a linear combination of velocity and acceleration
\beq
\label{peua}
\bfp_e=m{\bf u}-\frac{2}{3}\frac{e_0^2}{c^3}\,\bfa,
\eeq
{and we note that adopting it as an exact equation,  implicitly the definition of the electron mass has to be changed: it is the  coefficient of the term linear in electron velocity from the electron momentum.  This change  is required since the electron momentum does not coincide  with the kinetic term $m{\bf u}=mc\gamma\bfbe$ when  $\dot{\bfbe}\neq 0$.} In particular,  if the electron velocity vanishes,   the momentum $\bfp_e$ of the accelerated electron reduces to the field momentum (\ref{pfrest}), 
\beq
\label{peir}
\bfp_e=-\frac{2}{3}\frac{e_0^2}{c^3}\,\dot{\bfv},  \quad \texttt{if } \bfv=0.
\eeq
This result is rather unusual since  we have a momentum of the  electron  proportional to its acceleration, not to its velocity.

{

We mention other implications of the equations (\ref{peS}) and (\ref{peS0}), defining the components of Schott momentum. These equations can be put in the simple form 
\beq
\label{psmta}
p_S^0=-m\tau_0\,a^0,\quad \bfp_S=-m\tau_0\,\bfa,
\eeq
where $(a^0,\bfa)$ is the 4-acceleration (\ref{4acc}) and 
\beq
\tau_0\equiv\frac{2}{3}\frac{r_0}{c}=\frac{2}{3}\frac{e_0^2}{mc^3}
\eeq
is the time for light propagation on a distance equal to 2/3 of  the classical radius of the electron, $r_0$.  
{The proportionality between  the 4-vector $(p_S^0,\bfp_S)$ and 4-acceleration of the electron allows us to write a simple formula connecting  the 4-scalars  $\bfp_S^2-(p_S^0)^2$  and ${\cal P}_L$ [the power  (\ref{LP})] (both  proportional to the 4-scalar (\ref{a2a02})), 
\beq
\label{pspl}
\bfp_S^2-(p_S^0)^2=m\tau_0\,{\cal P}_L.  
\eeq
Using Eqs. (\ref{psmta}) and ({\ref{a0ba}) one can  express the Schott energy $W_S$  in terms  of electron velocity and Schott momentum 
\beq
\label{wshpsh}
W_S=cp_S^0=c\bfbe\cdot\bfp_S.
\eeq

\vspace{2ex}

An interesting consequence of Eqs. (\ref{pef2}) and (\ref{wet}) is a modified energy-momentum relation, which we first write in the form 
 \beq
 \label{wpal1}
 W_e^2=m^2c^4-c^2\left[\bfp_S^2-(\bfbe\cdot\bfp_S)^2\right]
 +c^2\,\bfp_e^2.
\eeq
For its demonstration one can start, for example, with the calculation of the difference $W_e^2-c^2\bfp_e^2$,  using Eqs. (\ref{pef2}) and  (\ref{wet}), then one applies  Eq. (\ref{wshpsh}). 
Transforming the middle term of Eq. (\ref{wpal1}) on the base of   Eqs. (\ref{pspl}) and (\ref{wshpsh})  one obtains the more compact form  
 \beq
 \label{wpal2}
 W_e^2=m^2c^4\left(1-\frac{\tau_0}{mc^2}\,{\cal P}_L\right)
 +c^2\,\bfp_e^2.
\eeq

The relation (\ref{wpal2}) replaces Eq. (\ref{wp}) and reduces to it when the electron is {\it instantaneously} free, i.e. $\dot{\bfbe}=0$. 
An immediate implication of this relation } is that for the same momentum, the accelerated electron has a lower energy than a free one - we note that the condition "same momentum" implies different velocities for the free and the accelerated electron.  The effect is very small since the ratio $\rho\equiv {\cal P}_L\,\tau_0/{mc^2}$ is usually much less than unity. {Even assuming, for example, that an electron would radiate  10\% from its initial energy of 1 GeV  in a time of 10 fs, the corresponding (average) power  ${\cal P}_L$ is $10^{16}$ Mev/s  and the ratio $\rho$ is about $1.23\cdot 10^{-7}$.
}

\vspace{2ex}
{Adding to the momentum and energy of the electron the corresponding quantities of the radiation, given by Eqs. (\ref{prad}) and (\ref{wrad}) and let aside up to now,  we obtain the total momentum and energy 
of the compound system, having the form  anticipated in Eq. (\ref{pwt0}). They are 
\beq
\label{totp}
\bfP=\bfp_e+\bfp_{rad}=mc\gamma\bfbe+\bfp_S+\bfp_{rad},
\eeq
and 
\beq
\label{totw}
W=W_e+W_{rad}=mc^2\gamma+W_S+W_{rad}.
\eeq

}
\vspace{2ex}

With the help of the 4-vectors velocity (\ref{4velo}) and acceleration (\ref{4acc}), the above equations for momentum and energy of the electron and of radiation can be easily transcribed in manifestly covariant form.  Using Eqs. (\ref{pef2}) and (\ref{wet}), the 4-momentum of the electron,  
$(W_e/c, \bfp_e)$, has the components 
\beq
\label{pemu}
p_e^\mu=mu^\mu+{ p}_S^\mu,
\eeq
where the last term is the Schott 4-momentum 
\beq
{ p}_S^\mu=-\frac{2}{3}\frac{e_0^2}{c^3}\, a^\mu.
\eeq
{
It is to be mentioned that the validity of Eq. (\ref{pemu}) appears here as unconditioned -  the previous demonstration made in \cite{titbo} requires the condition of vanishing  acceleration of the electron  for $t\rar -\infty$; this might be due to a  different method used in \cite{titbo} to find the electron momentum, based on the calculation of the bound momentum rate.
}

The quantities referring  to radiation,  $W_{rad}/c$ and $\bfp_{rad}$ form together the radiation 4-momentum  $p^\mu_{rad}$. Using Eqs. ({\ref{wrad}) and ({\ref{prad}), its 
components are 
\beq
p^\mu_{rad}=
-\frac{2}{3}\frac{e_0^2}{c^5}\int_{-\infty}^{\tau}\left(a^\nu a_\nu\right) u^\mu \,d\tau',
\eeq
where  the Lorentz  invariant 
$a^\nu a_\nu=(a^0)^2-\bfa^2$ is, up to the sign, the 4-scalar (\ref{a2a02}).
}
For the  4-momentum of the whole system (electron and  radiation) we then have 
\beq
\label{pmu}
P^\mu=p^\mu_{e}+p^\mu_{rad}=mu^\mu-\frac{2}{3}\frac{e_0^2}{c^3}\,{a^\mu}-\frac{2}{3}\frac{e_0^2}{c^5}\int_{-\infty}^{\tau}\left(a^\nu a_\nu\right) u^\mu \,d\tau'.
\eeq

}

\vspace{2ex}

In the following we refer to the equations of motion of the electron in the field of an external force $\bfF$. { They can be obtained equating the rate of the {\it total} momentum (\ref{totp}) with the external force  
\beq
\label{mequ}
\frac{d\bfP}{dt}=\bfF,
\eeq
relation 
written here for a system which is not a purely  mechanic one.
Taking into account the structure of $\bfP$, Eq. (\ref{mequ}) can be set  in the form
\beq
\label{meqr}
m\frac{d}{dt}(\gamma\bfv)=\bfF+\Rcb,\quad \Rcb=\Rcb_S+\Rcb_L,
\eeq
where  the total reaction force $\Rcb$ is composed from 
the Abraham-Lorentz force (self-force) 
\beq
\label{selff}
\Rcb_S\equiv -\frac{d\bfp_S}{dt}=\frac{2}{3}\frac{e_0^2}{c^3}\,\frac{d\bfa}{dt}
\eeq
and the radiative damping  force 
\beq
\label{rare}
\Rcb_L\equiv -\frac{d\bfp_{rad}}{dt}=-\frac{2}{3}\frac{e_0^2}{c^3}\,
\gamma^4\left[ {\dot{\bfbe}}^2+\gamma^2(\bfbe\cdot\dot{\bfbe})^2 \right] \bfv.
\eeq
Eq. (\ref{meqr}) can be regarded as an  equation for $\bfr_e(t)$ (of the third order, due to the form of $\Rcb_S$) or as forming, together with $d\bfr_e/dt={\bf u}/\gamma$, a system of equations of lower order for the variables $\bfr_e(t)$ and ${\bf u}(t)$.
It presents also interest to express  Eq. (\ref{mequ}) as an equation for the momentum $\bfp_e$, 
\beq
\label{meqrl}
\frac{d\bfp_{e}}{dt}=\bfF+\Rcb_L,
\eeq
the system of motion equations being completed with Eq. (\ref{peua}), transcribed as
\beq
\label{upa}
m\gamma\frac{d\bfr_e}{dt} =\bfp_e+\frac{2}{3}\frac{e_0^2}{c^3}\,\bfa.
\eeq

 The elementary work done by the force $\bfF$ is $\bfF\cdot d\bfr=\bfF\cdot \bfv\,dt$, where $d\bfr$ is the position change of the electron in the time interval $dt$. { One can justify separately the following equalities
\beq
\label{2eqs}
 c\bfbe\cdot\frac{d}{dt}\left(mc\gamma\bfbe\right)=\frac{d}{dt}\left(mc^2\gamma\right),\quad  c\bfbe\cdot\left(\frac{d\bfp_S}{dt}+\frac{d\bfp_{rad}}{dt}\right)=\frac{dW_S}{dt}+\frac{dW_{rad}}{dt}.
\eeq
 An elementary  calculation based on  Eqs. (\ref{mequ}), (\ref{totp}) and (\ref{totw}), and using the sum of the relations (\ref{2eqs}), gives for the  work per time unit }
\beq
\label{vf}
\bfv\cdot\bfF=c\bfbe\cdot\bfF=\frac{dW}{dt},
\eeq
expressing the energy conservation. Integrated over a finite time interval, it tells us that the corresponding work of the external force is equal to the sum of the energy lost by radiation  and of the variations of the kinetic energy and Schott energy  on that time interval.

The  equations (\ref{mequ}) and (\ref{vf}) can be grouped in manifestly covariant form,  
\beq
\frac{dP^\mu}{d\tau}=K^\mu, 
\eeq
where $K^\mu=(\gamma \bfbe\cdot\bfF, \gamma\bfF)$ is the 4-force.  Differentiating to $\tau$  the momentum (\ref{pmu}) one finds the explicit form of the  equation {of motion}, 
\beq
\label{ladk}
m\frac{du^\mu}{d\tau} -\frac{2}{3}\frac{e_0^2}{c^3}\,\frac{d^2u^\mu}{d\tau^2}-\frac{2}{3}\frac{e_0^2}{c^5}\,\left(a^\nu a_\nu\right) u^\mu=K^\mu,
\eeq
(by $u^\mu$ one understands $dx^\mu/d\tau$) reducing to LAD equation when the force $\bfF$ is the Lorentz force describing the interaction of the electron with an external electromagnetic field.

We finally note that Eq. (\ref{ladk}) (or its  spatial part (\ref{meqr}))  is not the most suitable starting point for applications. 
For the accurate numerical solving of the motion problem or for deriving practical approximations (the subject shall be developed  elsewhere, here we present just few remarks) it is more convenient to  use in place of  the LAD equation  an equivalent system of differential equations of lower order, objective that can be reached in several ways,
differing by the choice of the dependent variables. The choice of  the position $\bfr_e$ and the momentum $\bfp_e$ of the electron as unknowns (see Eqs. (\ref{meqrl}) and (\ref{upa})) is  attractive especially when it is combined with the  approximation method based on the smallness of RR effect.  Neglecting RR in a first stage, one obtains  an estimation for the acceleration of the electron $\bfa\approx {\gamma}\,\bfF/m$. When this approximation is used to evaluate the  term  $\Rcb_L$ in Eq.  (\ref{meqrl}), and the Schott momentum in Eq. (\ref{upa}), one get reduced-order  {\it  approximate} versions of these equations.  
For the case of an external electromagnetic force, the  equations so obtained  should be equivalent with the ones proposed in \cite{soko} (more precisely, with their spatial part), in the same order of approximation (first order of the parameter $\tau_0$). 
As also discussed in \cite{burt},  one has to be cautious with such kind of approximation - in particular,  Eq. (\ref{upa}) (with $\bfa$ replaced by ${\gamma}\,\bfF/m$)  could lead to superluminal electron velocities when used in conditions such RR is not a small perturbation of the system.

\section{Angular momentum}

\label{amo}

We consider first the simple example of an electron at rest up to the time moment $t=0$, accelerated on the time interval $0<t<T$ by an external force $\bfF$, and free again for $t>T$. For $t\gg T$ we would expect that the total angular momentum of the system (electron and emitted radiation) to be simply the sum of two terms: the angular momentum of the electron, $\bfr_e\times (m\gamma\bfv)$, and the angular momentum $\bfL_{rad}$ of radiation. In fact, as discussed below, due to the radiation reaction on the interval $(0,T)$ this intuitive feature is not confirmed in general.

We present  in the following the calculation and the results obtained for the total angular momentum of an accelerated electron and its field - the procedure used, similar to that of Sect. II, starts from the electromagnetic field of the electron.   
We take as definition for the density of the angular momentum of the electromagnetic field the quantity $\bfl=\bfr\times \bfg $, where $\bfg$ is the momentum density (\ref{gu}). 
Then the contribution of the above defined  region $\cal D$ (the region outside the sphere  of radius $\veps$, centered on the  retarded position $\bfr_e(t-\veps/c)$ of the electron)   to the angular momentum is 
\beq
\label{totld}
\bfL_{em}(\veps,t)\equiv \int_{{\cal D}}\bfr\times\bfg\,d\bfr.
\eeq
Replacing $\bfr=\bfr_e(t')+\bfR(t')$ {(see the Figure and Eq. (\ref{tret}))}, $\bfL_{em}(\veps,t)$ can be written as a sum of two terms, $\bfL_{em}(\veps,t)=\bfL_1(\veps,t)+\bfL_2(\veps,t)$, where
\beq
\label{l12}
\bfL_1(\veps,t)\equiv\int_{{\cal D}}\bfr_e{(t')}\times\bfg\,d\bfr, \quad \bfL_2(\veps,t)\equiv\int_{{\cal D}}\bfR{(t')}\times\bfg\,d\bfr.
\eeq
The calculation of  $\bfL_{1,2}$, similar with that corresponding to  the linear momentum, is also presented  in the Appendix. 
The results  are 
\beq
\label{l1eps}
\bfL_1(\veps,t)=\frac{2}{3}\frac{e_0^2}{c}\,\bfr_e(t_\veps)\times\frac{\gamma^2(t_\veps)\bfbe(t_\veps)}{\veps}+\frac{2}{3}\frac{e_0^2}{c^2}\int_{-\infty}^{t_\veps}
\gamma^4\left[ {\dot{\bfbe}}^2+\gamma^2(\bfbe\cdot\dot{\bfbe})^2 \right]\bfr_e\times \bfbe\,dt',
\eeq
and
\beq
\label{l2eps}
\bfL_2(\veps,t)=\frac{2}{3}\frac{e_0^2}{c}\int_{-\infty}^{t_\veps}
\gamma^2\,\bfbe\times \dot{\bfbe}\,dt'.
\eeq
We note that the first term of $\bfL_1$ is simply the product $\bfr_e(t_\veps)\times \bfp_f(\veps,t)$ {(see Eq. (\ref{pfat0}))}. 
In order to obtain the total angular momentum of the electron and its field we add the contribution of the bare electron, assumed to be  $\bfr_e(t_\veps)\times\overline{\bfp}_e(\veps,t)$, to the sum $\bfL_1(\veps,t)+\bfL_2(\veps,t)$ and take the limit $\veps\rightarrow 0$ using  the first relation (\ref{pef3}) and Eq. (\ref{pef2}). The final result can be conveniently expressed as

\beq
\label{angm}
\bfL=\bfL_e+\bfL_{rad}+\bfL_x,
\eeq
where
\beq
\bfL_e\equiv \bfr_e\times \bfp_e,\quad \bfL_{rad}\equiv 
\frac{2}{3}\frac{e_0^2}{c^2}\int_{-\infty}^{t}
\gamma^4\left[ {\dot{\bfbe}}^2+\gamma^2(\bfbe\cdot\dot{\bfbe})^2 \right]\left(\bfr_e\times \bfbe\right)\,dt',
\eeq
and
\beq
\label{lx}
\bfL_x\equiv \frac{2}{3}\frac{e_0^2}{c}\int_{-\infty}^{t}
\gamma^2\,\left(\bfbe\times \dot{\bfbe}\right)\,dt'.
\eeq
The term  $\bfL_e$ is the angular momentum of the accelerated electron; it differs from  the usual definition of the angular momentum of a particle by the supplementary term  $\bfr_e\times\bfp_S$. The next term,  $\bfL_{rad}$, equal to the integral of $\bfr_e(t')\times d\bfp_{rad}(t')$, where $d\bfp_{rad}(t')$ is the momentum taken by radiation in  $dt'$,  can be identified with the total angular momentum of the radiation emitted by the electron up to the time $t$.  Unlike  $\bfL_e$, $\bfL_{rad}$ and $\bfL_x$ depend  on the past motion of the electron. This feature is rather intriguing for $\bfL_x$, it raising an interpretation difficulty concerning the subsystem to which this term has to be attributed. {Not being a function of the actual state of the electron,  $\bfL_x$  is not an attribute of the electron.  The only possibility remained is to regard it as a property of the total system (electron and radiation), interpretation supported by the fact that  it originates from a mixed contribution of the fields.  }
We note that the presence of this term is necessary for the balance of the angular momentum. Indeed, defining conveniently 
\beq
\bfL_S\equiv \bfr_e\times \bfp_S+\bfL_x,
\eeq
and using under the integral for $\bfL_x$ the identity 
$$c\gamma^2\,\left(\bfbe\times \dot{\bfbe}\right)=\frac{d}{dt}\left[
\bfr_e\times\left(\gamma\frac{d}{dt}(\gamma\bfbe)\right)\right]-\bfr_e\times\frac{d}{dt}\left(\gamma\frac{d}{dt}(\gamma\bfbe)\right),
$$
and the expression  (\ref{peS}) of the Schott momentum, one obtains first  the relation
\beq
\bfL_S=\int_{-\infty}^{t}
\bfr_e\times\frac{d\bfp_S}{dt'}\,dt'.
\eeq
Then one can express Eq. (\ref{angm}) in the equivalent form 
\beq
\bfL= \bfr_e\times (m c\gamma\bfbe)+\int_{-\infty}^{t}
\bfr_e\times\frac{d\bfp_S}{dt'}\,dt'+
\int_{-\infty}^{t}
\bfr_e\times\frac{d\bfp_{rad}}{dt'}\,dt',
\eeq
 allowing to calculate easily the angular momentum rate  
\beq
\frac{d\bfL}{dt}=\bfr_e\times \frac{d}{dt}(m c\gamma\bfbe+\bfp_S+\bfp_{rad}).
\eeq
Using Eqs. (\ref{totp}) and (\ref{mequ}), we obtain the equality 
\beq
\label{amth}
\frac{d\bfL}{dt}=\bfr_e\times \bfF,
\eeq
which is the generalization of the angular momentum theorem to the actual system. 
{The compatibility of Eq. (\ref{amth})  with LAD equation agrees with  the findings of \cite{bhabha} (see  also Sect. \ref{intr}).}

We return here to the simple case mentioned at the beginning of the Section and analyze  the consequences of Eqs. (\ref{angm}) and (\ref{amth}) {for $t>T$, where}  {the total angular momentum $\bfL$ is conserved ($\bfF=0$).} 
{In fact, all the terms of Eq. (\ref{angm}) are constant since $\dot{\bfbe}=0$.}
Consequently, they have the values 
\beq
\bfL_e= \bfr_e(T)\times (mc\bfbe(T)),\quad \bfL_{rad}= 
\frac{2}{3}\frac{e_0^2}{c^2}\int_{0}^{T}
\gamma^4\left[ {\dot{\bfbe}}^2+\gamma^2(\bfbe\cdot\dot{\bfbe})^2 \right]\left(\bfr_e\times \bfbe\right)\,dt,
\eeq
and 
\beq
\label{lxsc}
\bfL_x\equiv \frac{2}{3}\frac{e_0^2}{c}\int_{0}^{T}
\gamma^2\,\left(\bfbe\times \dot{\bfbe}\right)\,dt.
\eeq
The presence of a constant non-vanishing $\bfL_x$, no matter how large is $t-T$, is puzzling since it involves the conclusion that a full separation of the electron  and its radiation is not possible. This can be better understood if we recall that  the spatial region contributing to $\bfL_x$ at a given $t>T$ is the region where both the velocity and the acceleration fields are non-vanishing (see the Appendix). This region, also wholly containing the radiation (whose properties are determined by the acceleration fields), is not a fixed one - the surfaces which bound it are the spherical surfaces of radii $c(t-T)$ and $ct$, centered  respectively on $\bfr_e(T)$ and $\bfr_e(0)$. 
The fields it contains can be seen as an electromagnetic perturbation of the velocity fields - this perturbation "kills" in front of it the Coulomb field and lets behind it the fields of an electron in free motion.

Assuming that the interpretation proposed for the angular momentum $\bfL_x$ is correct, we might ask ourselves how it can be observed. The direct observation of $\bfL_x$ may be difficult, but an indirect way,  based on the conservation of the total angular momentum, could be considered.  Referring again to the above example, in the case of an external force $\bfF$ of {\it central} type, the torque $\bfr_e\times \bfF$ is zero, and the conservation of $\bfL$ on $(0,T)$ imposes that 
$\bfL(t)=\bfL(0)=0$ for any $t>0$ (the electron is initially  at rest and no radiation is present). For the sake of simplicity, we assume the electron is left at rest also for $t>T$. In this case $\bfL_e=0$ and  the equality $\bfL_{rad}+\bfL_x=0$ implies that if one detects a non-vanishing angular moment of the radiation {(with the negligible perturbation of the radiation itself)},  the same property is shared by $\bfL_x=-\bfL_{rad}$. These considerations can be easily extended for non-central forces and any initial and final states of the electron, invoking in this case the theorem (\ref{amth}) (in its integrated version, on the corresponding time interval).

\section{Conclusions}

For a point electron in accelerated motion we analytically  integrate the densities of {
momentum,  energy and angular momentum} of its electromagnetic field, {in a spatial region excluding a vicinity of the electron and extended to infinity.}  {
For the case of momentum and energy,} after separating the quantities referring  to  radiation, the singularities of the contributions coming from the  fields of the point electron are compensated by those of the bare electron contributions  such as to produce finite results,  satisfying at the same time relativistic covariance criteria. 

We infer this way  expressions  for the momentum and energy of an  electron in arbitrary motion, differing from those of the free electron by the Schott terms.
The corresponding equations, (\ref{pef2}) and (\ref{wet}), imply a modification of the energy-momentum relation of the free electron by a term proportional to the Li\'{e}nard-Larmor power of the emitted radiation. 

The inferred formula for the total momentum of the compound system (electron and radiation emitted by it) allows to derive in a simple manner  the electromagnetic reaction force and to formulate the equation (or the system of equations) of motion of the electron. 
{
For} an electron in an external electromagnetic field, the  equation of motion coincides  with the Lorentz-Dirac equation. 

{
 Tho formula obtained for the total angular momentum (by adding to the quantity corresponding to the electromagnetic field that of the bare electron and by taking the limit of the point  electron)  contains, besides the expected terms for the electron and radiation, a supplementary term which  can be non-vanishing  after the end of the interaction with the external field. It is shown that its presence is necessary for a correct balance of the angular momentum.}

}

\begin{acknowledgments}

This work has been supported by the Project 29/2016 ELI RO,  financed by the Institute of Atomic Physics. The author warmly 
thanks  Antonino Di Piazza and Viorica Florescu for their pertinent comments and useful suggestions.

\end{acknowledgments}

\appendix*{

\label{fme}

\section{Momentum, energy and angular momentum of the field in the domain $\cal D$ }

We present here in some detail the calculation of the integrals giving the momentum, energy and angular momentum of the electromagnetic field in the region $\cal D$ chosen as described in Sect. \ref{mae} and Fig. 1. We refer first to momentum and energy, whose expressions are  
\beq
\label{gawa}
\bfG(\veps,t)= \int_{{\cal D}}\bfg\,d\bfr, \quad W(\veps,t)=\int_{{\cal D}}u\,d\bfr,
\eeq
where the densities $\bfg$ and $u$ are given by  Eq. (\ref{gu}), for fields described in (\ref{ebfields}). 
We conveniently separate the fields  in velocity  and acceleration  fields 
\beq
\label{ebfvam}
\bfE(\bfr,t)=\bfE_v(\bfr,t)+\bfE_a(\bfr,t),\quad \bfB(\bfr,t)=\bfB_v(\bfr,t)+\bfB_a(\bfr,t), 
\eeq
where
\beq
\bfE_v(\bfr,t)\equiv \frac{e}{4\pi\varepsilon_0}\frac{1-\beta^2}{R^2\varkappa^3}\,(\bfn-\bfbe), \quad \bfB_v(\bfr,t)\equiv\frac{1}{c}\,\bfn\times \bfE_v(\bfr,t),
\eeq
and 
\beq
\bfE_a(\bfr,t)\equiv \frac{e}{4\pi\varepsilon_0 c}\frac{1}{R\varkappa^3}\,\bfn\times\left[(\bfn-\bfbe)\times\dot{\bfbe}\right], \quad \bfB_a(\bfr,t)\equiv\frac{1}{c}\,\bfn\times \bfE_a(\bfr,t).
\eeq
Then the densities $\bfg$ and $u$, quadratic in fields,  can be written as 
\beq
\bfg=\bfg_v+\bfg_x+\bfg_a,\quad u=u_v+u_x+u_a, 
\eeq
where the indexes "$v$" and "$a$" denote, respectively, contributions of velocity and acceleration fields, {while "$x$" is index for the mixed terms,} expressed with both types of fields.

The spatial integration on $\cal D$ 
can be performed changing first the integration variables from the coordinates $\bfr$ to the relative coordinates $\bfR=\bfr-\bfr_e(t')$, with $t'=t-R/c$. This is justified since, for a given actual time $t$, the time moment $t'$ is uniquely determined by the position $\bfr$. The uniqueness of $t'$ and of $\bfr_e(t')$ involves that of $\bfR$ for each $\bfr$.  Using the inverse transformation,  $\bfr=\bfR+\bfr_e(t-R/c)$,  one easily calculates the Jacobian $(\partial \bfr/\partial\bfR)$, this being $\varkappa=1-\bfn\cdot\bfbe$, where $\bfn$ is the unit vector along $\bfR$.  Then passing  to the spherical coordinates of $\bfR$, we get the integration "recipe" 
\beq
\label{inre}
 \int_{{\cal D}}\ldots\,d\bfr =\int_{R>\veps}\ldots \varkappa\,d\bfR
 =\int_{\veps}^{\infty}R^2\left[\int \varkappa \ldots 
 d\Omega_{\bfn}\right]\, dR. 
\eeq

The  angular integrals over the  direction $\bfn$, involved in   the calculation of integrals (\ref{gawa}), can  all be  expressed by the simpler ones,
\beq
J^{(m)}(\beta)\equiv \int \frac{1}{\varkappa^{m+1}}\, d\Omega_{\bfn}=
\int_0^\pi \, \int_0^{2\pi}\frac{1}{(1-\bfn\cdot\bfbe)^{m+1}}\,\sin\vartheta\, d\vartheta\,d\varphi,\quad m=1,2,\ldots,
\eeq
and their partial derivatives to the components of the 3-vector $\bfbe$. 
For the scalar $J^{(m)}(\beta)$, simply to calculate with the choice of polar axis along $\bfbe$, one obtains the result 
\beq
J^{(m)}(\beta)=\frac{2\pi}{m\beta}\left[\frac{1}{(1-\beta)^m}-\frac{1}{(1+\beta)^m}\right].
\eeq

We present first the results for the angular integrals corresponding to the momentum $\bfG$,
\beq
\label{aigv}
\int \varkappa \,\bfg_v\, d\Omega_{\bfn}=\frac{2}{3}\,\frac{e_0^2}{cR^4}\,\gamma^2\bfbe,\quad e_0\equiv\frac{e}{\sqrt{4\pi\varepsilon_0}},
\eeq
\beq
\label{aigm}
\int \varkappa \,\bfg_x\, d\Omega_{\bfn}=\frac{2}{3}\,\frac{e_0^2}{c^2R^3}\,\frac{d}{dt'}(\gamma^2\bfbe),
\eeq
and 
\beq
\label{aiga}
\int \varkappa \,\bfg_a\, d\Omega_{\bfn}=\frac{2}{3}\,\frac{e_0^2}{c^3R^2}\,\gamma^4  \left[ {\dot{\bfbe}}^2+\gamma^2(\bfbe\cdot\dot{\bfbe})^2 \right] \bfbe.
\eeq
 The  next relations give  the angular integrals for the case of energy $W$, 
\beq
\int \varkappa \,u_v\, d\Omega_{\bfn}=\frac{2}{3}\,{e_0^2}\,\frac{\gamma^2-1/4}{R^4},
\eeq
\beq
\int \varkappa \,u_x\, d\Omega_{\bfn}=\frac{2}{3}\,\frac{e_0^2}{cR^3}\,\frac{d}{dt'}(\gamma^2),
\eeq
and 
\beq
\int \varkappa \,u_a\, d\Omega_{\bfn}=\frac{2}{3}\,\frac{e_0^2}{c^2R^2}\,\gamma^4  \left[ {\dot{\bfbe}}^2+\gamma^2(\bfbe\cdot\dot{\bfbe})^2 \right].
\eeq
We recall that the time argument of the above expressions is $t'=t-R/c$. Then, for a fixed $t$ and arbitrary function $f(t')$, we have $df/dt'=-c \, df/dR$, a relation useful for the integration  over $R$. 
Using the recipe (\ref{inre}) for the sum $\bfg_v+\bfg_x$ and the relations (\ref{aigv}) and (\ref{aigm}), we have 
\beq
\int_{{\cal D}}(\bfg_v+\bfg_x)\,d\bfr=\frac{2}{3}\,\frac{e_0^2}{c}\,\int_{\veps}^{\infty}
\left[
\frac{\gamma^2\bfbe}{R^2}+\frac{1}{cR}\frac{d}{dt'}(\gamma^2\bfbe)
\right]\,dR, 
\eeq
where the function under the integral sign is the derivative to $R$  of $(-\gamma^2\bfbe/R)$. Consequently, 
\beq
\label{gvm}
\int_{{\cal D}}(\bfg_v+\bfg_x)\,d\bfr=\frac{2}{3}\,\frac{e_0^2}{c}\,\frac{\gamma^2(t_{\veps})\bfbe(t_{\veps})}{\veps},
\eeq
with $t_\veps=t-\veps/c$.

The same  recipe (\ref{inre}), applied for  $\bfg_a$ with the variable change from $R$ to $t'=t-R/c$, furnishes the result 
\beq
\label{ga}
\int_{{\cal D}}\bfg_a\,d\bfr=
\frac{2}{3}\,\frac{e_0^2}{c^2}\,\int_{-\infty}^{t_{\veps}}\gamma^4  \left[ {\dot{\bfbe}}^2+\gamma^2(\bfbe\cdot\dot{\bfbe})^2 \right] \bfbe\, dt'.
\eeq

A similar calculation performed for the total energy of the region $\cal D$ leads to the following results

\beq
\label{uvm}
\int_{{\cal D}}(u_v+u_x)\,d\bfr=\frac{2}{3}\,{e_0^2}\,
\frac{\gamma^2(t_{\veps})-1/4}{\veps},
\eeq
and
\beq
\label{ua}
\int_{{\cal D}}u_a\,d\bfr=
\frac{2}{3}\,\frac{e_0^2}{c}\,\int_{-\infty}^{t_{\veps}}\gamma^4  \left[ {\dot{\bfbe}}^2+\gamma^2(\bfbe\cdot\dot{\bfbe})^2 \right]\, dt'.
\eeq
}

Using  Eqs.  (\ref{gvm}) and (\ref{ga}), then  (\ref{uvm}) and (\ref{ua}), one gets the final expressions, {reproduced in Eqs. (\ref{glam2}) and (\ref{wd})}, for the quantities in Eq. (\ref{gawa}).

\vspace{2ex}

In the case of the angular momentum (\ref{totld}), the integrals we have to evaluate are given by Eq. (\ref{l12}). Using the recipe (\ref{inre}), we transcribe them here in the form 
\beq
\bfL_1(\veps,t)=\int_{\veps}^{\infty}R^2\left[\int \varkappa (\bfr_e\times \bfg)\,d\Omega_{\bfn}\right]\, dR, \quad 
\bfL_2(\veps,t)=\int_{\veps}^{\infty}R^2\left[\int \varkappa (\bfR\times \bfg)\,d\Omega_{\bfn}\right]\, dR.
\eeq
We  note that  the calculation of $\bfL_1$ is very simple since the vector $\bfr_e=\bfr_e(t-R/c)$ does not depend on the direction $\bfn$ of $\bfR$. Then 
\beq
\bfL_1(\veps,t)= \int_{\veps}^{\infty}R^2\bfr_e\times\left[\int  \varkappa ( \bfg_v+\bfg_x+\bfg_a)\,d\Omega_{\bfn}\right]\, dR.
\eeq
The integrals over directions are already met - see Eqs. (\ref{aigv})-({\ref{aiga}). Performing their sum and the integration in $R$ one obtains the formula (\ref{l1eps}).

For $\bfL_2$ we have, using $\bfR=R\bfn$, 
\beq
\bfL_2(\veps,t)=\int_{\veps}^{\infty}R^3\left[\int \varkappa (\bfn\times \bfg)\,d\Omega_{\bfn}\right]\, dR.
\eeq
The calculation of the angular integral is simplified if one 
use the equalities $\bfn\times \bfg=-\frac{\veps_0}{c}(\bfn\cdot\bfE)(\bfn\times\bfE)=-\frac{\veps_0}{c}(\bfn\cdot\bfE_v)(\bfn\times\bfE)$, implying that
$$\int \varkappa (\bfn\times \bfg)\,d\Omega_{\bfn}=-\frac{\veps_0}{c}
\int \varkappa (\bfn\cdot\bfE_v)(\bfn\times\bfE_a)\,d\Omega_{\bfn}.$$
The last equality shows that the field contribution to $\bfL_2$ is of a mixed type. The effective calculation yields  first the result 
\beq
\int \varkappa (\bfn\times \bfg)\,d\Omega_{\bfn}=
 \frac{2}{3}\frac{e_0^2}{c}\,
\gamma^2\,\left(\bfbe\times \dot{\bfbe}\right), 
\eeq
then, after the variable change to $t'=t-R/c$, the relation (\ref{l2eps}).
\end{document}